\documentclass[reprint,superscriptaddress,amsmath,amssymb,aps,pre]{revtex4-1}

\usepackage{graphicx}
\usepackage{dcolumn}
\usepackage{bm}
\usepackage{color}
\usepackage{bbold}
\usepackage{soul}
\newcommand{\newc}{\newcommand}
\newc{\beq}{\begin{equation}}
\newc{\eeq}{\end{equation}}
\newc{\kt}{\rangle}
\newc{\br}{\langle}
\newc{\beqa}{\begin{eqnarray}}
\newc{\eeqa}{\end{eqnarray}}
\newc{\longra}{\longrightarrow}

\providecommand*{\ue}{\text{e}}
\providecommand*{\ui}{\text{i}}

\providecommand*{\Trace}{\text{Tr}}

\makeatletter
\let\Hy@backout\@gobble
\makeatother

\begin{document}

\title{Universal scaling of spectral fluctuation transitions for interacting
       chaotic systems}

\author{Shashi C. L. Srivastava}
\altaddress[Permanent address: ]{Variable Energy Cyclotron Centre, Kolkata 700064, India.}
\affiliation{Max-Planck-Institut f\"ur Physik komplexer Systeme, N\"othnitzer
Stra\ss{}e 38, 01187 Dresden, Germany}

\author{Steven Tomsovic}
\altaddress[Permanent address: ]{Department of Physics and Astronomy,
            Washington State University, Pullman, WA~99164-2814}
\affiliation{Technische Universit\"at Dresden, Institut f\"ur Theoretische
             Physik and Center for Dynamics, 01062 Dresden, Germany}

\author{Arul Lakshminarayan}
\altaddress[Permanent address: ]{Department of Physics,
            Indian Institute of Technology Madras, Chennai, India~600036}
\affiliation{Max-Planck-Institut f\"ur Physik komplexer Systeme,
             N\"othnitzer Stra\ss{}e 38, 01187 Dresden, Germany}

\author{Roland Ketzmerick}
\affiliation{Max-Planck-Institut f\"ur Physik komplexer Systeme, N\"othnitzer
Stra\ss{}e 38, 01187 Dresden, Germany}
\affiliation{Technische Universit\"at Dresden, Institut f\"ur Theoretische
             Physik and Center for Dynamics, 01062 Dresden, Germany}

\author{Arnd B\"acker}
\affiliation{Max-Planck-Institut f\"ur Physik komplexer Systeme, N\"othnitzer
Stra\ss{}e 38, 01187 Dresden, Germany}
\affiliation{Technische Universit\"at Dresden, Institut f\"ur Theoretische
             Physik and Center for Dynamics, 01062 Dresden, Germany}

\date{\today}

\begin{abstract}
  The spectral properties of interacting strongly chaotic systems
  are investigated for growing interaction strength.
  A very sensitive transition from Poisson statistics to
  that of random matrix theory is found.
  We introduce a new random matrix ensemble modeling this
  dynamical symmetry breaking transition
  which turns out to be universal and depends on a single scaling
  parameter only.
  Coupled kicked rotors, a dynamical systems paradigm
  for such transitions, are compared with this ensemble
  and excellent agreement is found for the nearest-neighbor-spacing distribution.
  It turns out that this transition is described quite accurately using perturbation theory.
\end{abstract}

\pacs{PACS here}

\maketitle

Quantization of fully chaotic systems is known to lead to the spectral
fluctuations of random matrix theory (RMT)~\cite{Bohigas84}
and to exhibit energy level
repulsion~\cite{PorterBook}.  More generally, even non-integrable
models without apparent classical limits, such as spin systems, can also show
such features~\cite{Pals94,Kudo04}. In contrast, integrable systems generally
follow Poisson statistics, which are devoid of level
repulsion~\cite{Berry77b}. It is also well understood that combining spectra of
different irreducible representations tends toward Poisson statistics in the
limit of superposing many sequences~\cite{Pandey79,Brody81}.  An instance where
this occurs is for the spectra of two separable, but individually chaotic
systems.  Whereas each subsystem possesses RMT fluctuations, the full spectrum
tends to Poisson fluctuations in the large dimensionality limit~\cite{Tkocz12}.

This begs the question of what happens to spectral fluctuations
if such separable, but quantum chaotic
subsystems, interact. There are many motivations for studying such systems.
For example, they may be of direct physical interest, such as
conduction electrons in chaotic quantum dots interacting through a
screened Coulomb potential~\cite{Jalabertsp}.
Another motivation derives from quantum information theory where
the development of entanglement is of particular
importance~\cite{Lakshminarayan14}. Two quantum spin
chains with RMT spectral fluctuations coupled in a ladder configuration is
a many-body situation where such transitions are possible as well.

Typically, the interaction between two subsystems leads
to entanglement and paves the way for RMT fluctuations of the combined system.
This Poisson-to-RMT transition can be viewed as a dynamical symmetry
breaking in analogy to fundamental symmetry breaking; the first exact RMT solution to a symmetry
breaking problem involved time-reversal invariance~\cite{Pandey82}.  For
modeling a particular dynamical system, it is important to connect dynamical
system parameters with the abstract transition parameter built into an RMT
transition ensemble, and to confirm that this correctly describes the
dynamical system's statistical properties.   For example, partial transport barriers 
often arise classically and an RMT ensemble can correctly model their
effects, but only if they have been accounted for their respective fluxes, relative phase space volumes, and 
tendency to couple locally~\cite{Bohigas93,Michler12}.

There are a couple of other possible cases of a Poisson-to-RMT transition, a
metal-insulator transition where states transition from localized to
extended~\cite{Evers08}, and the perturbation of an
integrable dynamical system~\cite{Berry84,Bohigas93} that renders it chaotic
for sufficient perturbation strength.  In neither of these possibilities is
there a simple globally coupled Poisson-to-RMT ensemble.  Various
complications, such as the metal-insulator transition, the
KAM theorem regarding the survivability of tori, and
partial transport barriers, all prevent this.

In this paper we address how strong the interactions must be between the subsystems in
order to recover the spectral fluctuations of fully chaotic
systems, and how the Poisson-to-RMT transition occurs as the interaction 
magnitude varies from non-interacting to strongly interacting particles cases.
We show that the interplay between the coupling and an effective
Planck constant gives rise to a dimensionless scaling transition
parameter. This is obtained in much the same way as
found in the context of a global symmetry breaking in
RMT~\cite{French88a,Wigner67,Bohigas93,Bohigas95,Cerruti03,Guhr96}.
There, arbitrarily small couplings, but uniform everywhere, lead discontinuously to
level repulsion~\cite{Pandey81}.  If scaled properly, the entire transition is universal 
and predicted well by a perturbation theory.

\emph{Interacting systems.}---
To study the Poisson-to-RMT transition, consider 
the unitary Floquet operator, or kicked version, of generic
bipartite systems described by
\begin{equation}
\label{eq1}
{\cal U}=(U_1 \otimes
U_2)\, U_{12}
\end{equation}
where the $U_j$ are the subsystem Floquet operators and $U_{12}$ describes the interaction.  An example is the Hamiltonian
\begin{equation} \label{eq:hamiltonian}
 H = \frac{1}{2}(p_1^2+p_2^2) +
     \bigl[V_1(q_1)+V_2(q_2)+ b V_{12}(q_1,q_2)\bigr] \,\delta_t
\end{equation}
where $\delta_t= \sum_{n=-\infty}^{\infty} \delta(t-n)$ is a periodic train of
kicks, and a unit time has been chosen as the kicking period. The propagator
connecting states separated by one kick follows from $U_j=\exp[-i
p_j^2/(2 \hbar)] \exp(-i V_j/\hbar)$ ($j=1,2$) and $U_{12}=\exp(-i b V_{12}/\hbar)$.  The
classcial limit is a 4-dimensional symplectic map $(q_j,p_j) \mapsto
(q'_j,p'_j) =(q_j+p_j', p_j - \partial V_j/\partial q_j - b~\partial
V_{12}/\partial q_j)$, connecting the state of the system
immediately prior to consecutive kicks.

Perhaps the simplest paradigm for our purposes is that of coupled kicked
rotors~\cite{Froeschle72}.  They have been realized in experiments
on cold atoms~\cite{Gadway13}, can be made strongly chaotic, and the
interaction strength continuously varied.  The most elementary case is for two
interacting rotors with the single particle potentials $V_j=K_j \cos(2 \pi
q_j)/4 \pi^2$ and interaction
\begin{equation}
 V_{12}=\frac{1}{4 \pi^2}\, \cos[2 \pi(q_1+q_2)].
 \label{eq:intpot}
\end{equation}
The unit periodicity in the angle variables
$q_j$ is extended here to the momenta $p_j$ so that the phase space is
a 4-dimensional torus. If the kicking strengths, $\{K_1, K_2\}$, here $\{9,10\}$ respectively,
are each chosen sufficiently large, 
the individual maps are strongly chaotic with a Lyapunov
exponent of $\approx \ln (K_j/2)$~\cite{Chirikov79}.  The interaction strength
is tuned by the parameter $b$.

The quantum mechanics on a torus phase space of a single rotor 
gives rise to a finite Hilbert space of dimension $N$. The effective
Planck constant is $h=1/N$.  Thus $U_1$
and $U_2$ are $N$ dimensional unitary operators on their respective spaces
${\cal H}^N_1$ and ${\cal H}^N_2$ whereas the interaction $U_{12}$ is a $N^2$
dimensional unitary operator on the tensor product space
${\cal H}^N_1 \otimes {\cal H}^N_2$.
The quantized 4-dimensional map is given by
Eq.~(\ref{eq1}).
Such coupled quantum maps have been studied in different
contexts \cite{Lakshminarayan01,Richter14} where more details can be
found.  Here we are interested in the statistics of the eigenphases $\varphi_n$
defined by ${\cal U} |\psi_n\rangle = \ue^{\ui \varphi_n}  |\psi_n\rangle$.
The boundary conditions are chosen 
to break both parity and time-reversal symmetries.  Note that  
coupled kicked rotors with a different interaction term have also been 
studied on the cylinder~\cite{Shepelyansky94}.

\emph{RMT transition ensemble.}--- The first task is to construct the RMT transition ensemble 
associated with Eq.~(\ref{eq1}),
\begin{equation}
\label{eq:mod}
  {\cal U}_{\text{RMT}}(\epsilon) =
      \left( U^{(1)}_{\text{CUE}} \otimes U^{(2)}_{\text{CUE}}\right)
      \,  U_{12}(\epsilon)
\end{equation}
where the tensor product is taken of two independently chosen circular unitary
ensembles (CUE) of dimension $N$ representing the individual strongly chaotic
rotors, and $U_{12}(\epsilon)$ a diagonal unitary matrix in the resulting $N^2$
dimensional space representing the coupling.  Its diagonal elements are taken as $\exp(2 \pi i \epsilon \, \xi_{n_1n_2})$
where $\xi_{n_1n_2}$ are independent random variables that are uniform on the
interval $(-1/2,1/2]$, $\epsilon$ is a real number, and $1\le n_1,n_2 \le N$
label subsystem bases.  Any special structure of the rotor's $U_{12}$ is ignored because it turns out 
to be irrelevant for this study.  A limiting case of this ensemble has been studied previously~\cite{Lakshminarayan14}, wherein the entangling power of ${\cal U}_{\text{RMT}}(\epsilon=1)$ was found analytically. However, its spectral
features for general $\epsilon$ are not yet explored.  

For simplicity, the only statistical quantity considered in this paper
is the nearest neighbor spacing (NNS) distribution $P(s)$
of the eigenphases $\varphi_n$.
For $\epsilon=0$, ${\cal U}_{\text{RMT}}$ is the
direct (tensor) product of two independent CUE matrices.
In this case $P(s)$ has been shown analytically
as $N \to \infty$ to approach the Poissonian result
of $\exp (-s)$, where $s$ is the spacing in terms of unit mean~\cite{Tkocz12}.  The case $\epsilon=1$ represents 
strong coupling and CUE behavior. 

\begin{figure}[b]
\includegraphics[scale=1]{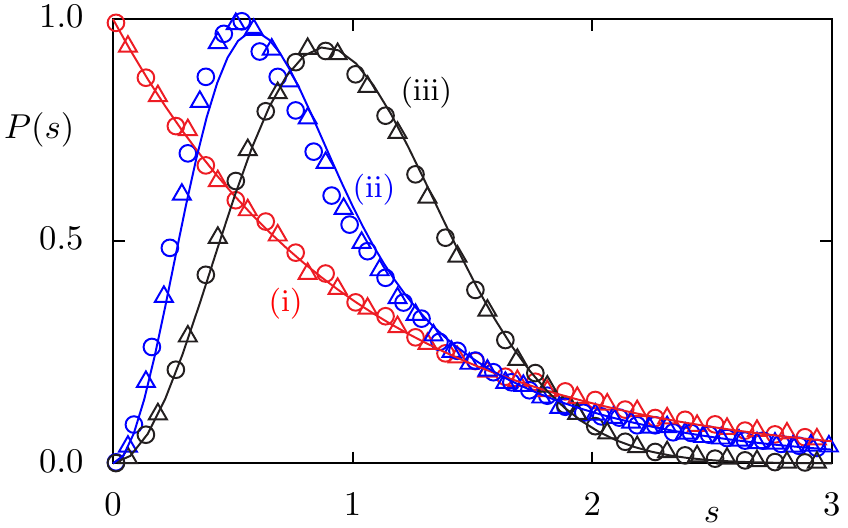}
\caption{\label{fig:gamma}(Color online)
  Nearest neighbor level-spacing distribution $P(s)$
  for two coupled rotors (circles), and the RMT model (triangles).
  (i) For $b=0$, it is
  Poissonian (red line) as is the $\epsilon=0$ RMT ensemble.
  (ii) For $b = 0.0019$, the spacing distribution is intermediate and
  agrees well with the corresponding RMT ensemble with $\epsilon = 0.012$
  and the result of a perturbation theory (blue line) for
  $\Lambda=0.1158$.
  (iii) For $b=0.008$, the spacing distribution agrees well with both the RMT ensemble ($\epsilon=0.05$) and the CUE.
For all cases the dimensionality of the unitary matrix is $N^2=10^4$.}
\end{figure}

\emph{Transition of the NNS distribution.}---
For non-interacting rotors,
the Poisson result is also expected to hold because 
the separability leads to spectral sequences that are composed of
superpositions of sequences.
For interacting rotors, the relation between $\epsilon$ of 
Eq.~(\ref{eq:mod}) and the $b$ of the coupled rotors
is needed.  As explained below, 
\begin{equation} \label{eq:epb1}
  \epsilon = \sqrt{\frac{3}{8\pi^4}} \,  N b
\end{equation}
for small values of $\epsilon$ and $Nb$, and $N\gg1$.
In practice this approximation is good for the entire transition
even if $N$ is only moderately large.

Figure \ref{fig:gamma} shows a comparison of the
results for the RMT transition ensemble and two coupled rotors
using Eq.~(\ref{eq:epb1}). This demonstrates that the ensemble
captures the transition in statistics from Poisson-like towards global CUE
perfectly well. CUE statistics already arise for small values
of $b$, which reflects the sensitivity of the transition.

\emph{Universal scaling.}---
The second task is to identify a universal scaling parameter, if it exists, and
test whether it captures the proper transition scale.
In other instances of symmetry
breaking~\cite{French88a,Bohigas93,Bohigas95,Cerruti03}, it
emerged using perturbation theory where it was given by
$\Lambda= v^2/D^2$, with $v^2$ the mean square off-diagonal matrix element
in the unperturbed system's diagonal representation and $D$ its mean
level spacing.  If $\Lambda \sim r$, where $r$ is the effective range of the
statistics (for the NNS, $r\sim 1$,) the transition is nearing completion.  It
must be emphasized that the interaction is assumed to be generic,
has no special symmetries and is entangling.
Ideally $v^2$ is calculated only from those
matrix elements that are responsible for the spectral transition, but averaging
over a global set of all matrix elements
gives the identical result if there is no secular structure,
e.g.\ like bandedness or special correlations \cite{French88b}.

For an $N^2$ dimensional unitary matrix, necessarily $D=2\pi/N^2$.  That leaves
$v^2$ to calculate.  The uncoupled part of the transition ensemble is diagonalized by the
direct product of two independent $N$ dimensional unitary matrices, say $u$ and
$w$.  An off-diagonal element of the interaction operator in the unperturbed
basis is $z_{kl;k'l'}=\sum_{m,n} u_{km} u_{k'm}^{*} w_{ln} w_{l'n}^{*}
(U_{12})_{mn}$, where $(k,l)\neq (k',l')$, and $(U_{12})_{mn}=\langle
mn|U_{12}|mn\rangle$ is a diagonal matrix element due to the interaction.  All
indices refer to subsystems and therefore lie in $[1, N]$. The off-diagonal
elements are restricted to $k \neq k'$ and $l \neq l'$, as the
energies of neighboring states after the tensor product must be different in
both.

The required average is that of $|z_{kl;k'l'}|^2$, as $u$ and $w$ are
independently chosen according to the Haar measure on $U(N)$,
the group of unitary
matrices in the subsystem spaces. This is done exactly using known results
for the average of $u_{km} u_{k'm' } u^*_{km'} u^*_{k'm'}$, which 
is $(\delta_{mm'}-1/N)/(N^2-1)$ when $k \neq k'$ (for example
see~\cite{Puchala11}), and leads to (for any $N$)
\begin{equation}
\begin{split}
& \Lambda=\dfrac{N^6}{4 \pi^2(N^2-1)^2} \times \\
& \left(1+ \left|\frac{1}{N^2}
  \Trace U_{12}\right|^2  - \frac{1}{N^3}\,
  \left[ \|U^{(1)} \|^2  + \|U^{(2)} \|^2 \right] \right).
\end{split}
\label{eq:lambda1}
\end{equation}
Here $ U^{(1)}_{ii}=\sum_{k}(U_{12})_{ik}$, $
U^{(2)}_{kk}=\sum_{i}(U_{12})_{ik}$ are partially traced (still diagonal)
interaction operators, which are in general not unitary, and
$\|X\|^2=\Trace(XX^{\dagger})$ is the Hilbert-Schmidt norm.  
Performing the exact ensemble average over $U_{12}(\epsilon)$ gives
\begin{equation}
\Lambda= \frac{N^4}{4\pi^2(N+1)^2}\left[1-\frac{\sin^2(\pi \epsilon)}{\pi^2 \epsilon^2}\right]  \approx \frac{N^2 \epsilon^2}{12 } ,
\label{eq:lambdarmt}
\end{equation}
where for the last approximation $N \gg 1$ and $\epsilon \ll 1$
have been used.  Note that related averages involving Haar measures over product spaces
have been obtained recently~\cite{Puchala12}.

It is important to recognize that Eq.~(\ref{eq:lambda1}), in the large-$N$ limit, applies to 
individual chaotic dynamical systems, and hence the coupled rotors.  The eigenbasis of uncoupled rotors 
behaves like that of a CUE member, and in the large-$N$ limit has 
the same statistical behavior as the ensemble.  Thus, applying Eq.~(\ref{eq:lambda1}) to the coupled rotors gives
\begin{equation} 
\label{eq:lambda-stdmap}
 \Lambda = \frac{N^2}{4\pi^2} \left(1-J_0^2(Nb/2 \pi)\right)
         \approx \frac{N^4 b^2}{32\pi^4},
\end{equation}
where $Nb \ll 1$ in the right form.  Using the approximations in Eqs.~(\ref{eq:lambdarmt},\ref{eq:lambda-stdmap}) generates Eq.~(\ref{eq:epb1}).  It can be checked that this $\Lambda$
correctly vanishes if $V_{12}$ is a function of either
coordinate alone, and thereby not entangling. However, it does not
vanish for general separable potentials, and hence the assumption that
{\it the interaction is entangling} is necessary.

\begin{figure}[b]
\includegraphics[scale=1.0]{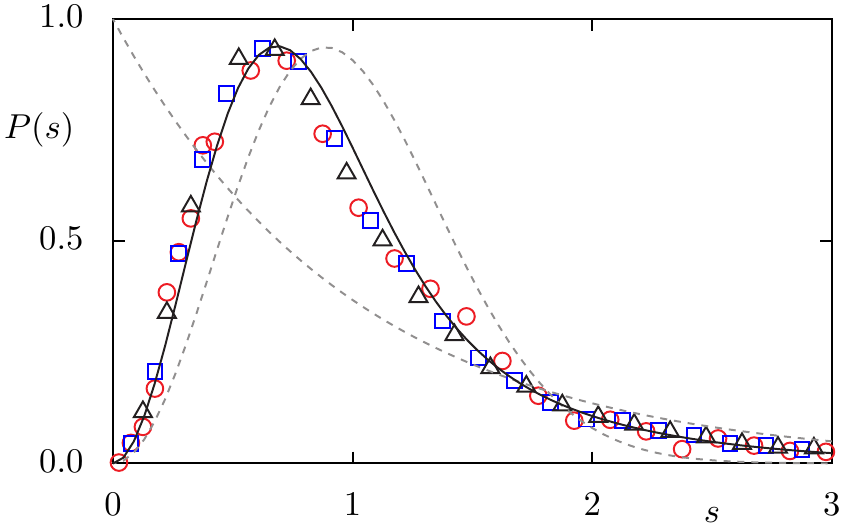}
\caption{(Color online)
  \label{fig:n2b} Universal scaling of the level-spacing distribution
  $P(s)$ for two coupled rotors.  Three results
  (triangles, squares, circles)  are shown for
  $(b,N)=(0.0025,100)$, $(0.01,50)$, and $(0.04,25)$. They all correspond to
  the same transition parameter $\Lambda \approx 0.2$. The solid
  curve comes from a perturbation theory described in the text
  and the dashed curves are the Poisson and CUE distribution, respectively.}
\end{figure}

There are two significant consequences of
Eqs.~(\ref{eq:lambdarmt},\ref{eq:lambda-stdmap}).  The first is that
there is a universal transition governed by the scaling parameter
$\Lambda \propto \epsilon^2 N^2 \propto b^2 N^4$.  Indeed, this is
well verified in Fig.~\ref{fig:n2b}.  Varying $b$ and $N$, but holding
$b^2 N^4$ fixed generates precisely the same statistical fluctuations.
Given that RMT ensemble statistics match those of the coupled rotors,
Fig.~\ref{fig:gamma}, the transition of Eq.~(\ref{eq:mod}) is the same for fixed $\epsilon^2 N^2$; this was also verified, but left out of Fig.~\ref{fig:n2b} to avoid clutter.  Secondly, the transition is very sensitive to the interaction strength.  The further a system is in the short wavelength limit, i.e. $\hbar \rightarrow 0$, the smaller the interaction strength needed to drive the transition to completion.  This is seen in the small values of $b,\epsilon$ chosen in Figs.~\ref{fig:gamma},\ref{fig:n2b}.  In fact, the transition is discontinuous in the interaction strength as $N\rightarrow \infty$~\cite{Pandey81}.

For Hamiltonians of the Eq.~(\ref{eq:hamiltonian})-type and
perturbative interactions (second order in $Nb$),
Eq.~(\ref{eq:lambda1}) simplifies to
\begin{equation}
\Lambda=N^4 b^2\left[ \langle V_{12}^2\rangle_{12} + \langle V_{12}\rangle_{12}^2 - \langle \langle V_{12}\rangle_1^2 \rangle_2- \langle \langle V_{12}\rangle_2^2 \rangle_1 \right].
\label{eq:lambda2}
\end{equation}
Here $\langle X \rangle_j = \Trace_j X/N$,
$\langle X \rangle_{12} = \Trace_{12} X/N^2$.
For the coupled rotors, with the interaction
as in Eq.~(\ref{eq:intpot}), the last three terms vanish in the large $N$ limit
as the integral over $q_1$ or $q_2$ is zero. Under such conditions, $\Lambda
\approx N^4 b^2\, \langle V_{12}^2 \rangle$,
where $\langle X\rangle = \int_0^1\int_0^1 dq_1 dq_2 X(q_1,q_2)$ and $X$ is a
classical function of the two coordinates. The interaction is
assumed to be diagonal in the position basis for the continuum approximation.

\emph{Perturbation theory.}--- A perturbative theory can be derived based on
previously developed techniques~\cite{French88a,Tomsovicthesis}.
The first step is to consider a very large dimensional
case and the perturbation expansion for the eigenvalues to second order.  The
first order term is diagonal and shifts the energies around randomly, and thus
does not alter the Poisson nature of the spectral statistics.  Amongst the
second order terms, terms connecting lower energies push a level up and those
above push down.  Overall, a level's motion fluctuates, but remains in the same
neighborhood.  However, when calculating the NNS spacing, there is a common
term, which enters opposite in sign for the two levels, and always pushes them
apart.  To a rough approximation, the bulk of the perturbation terms keep the
mean spacing constant, but the exceptional term introduces level repulsion.

\begin{figure}[b]
\includegraphics[scale=1]{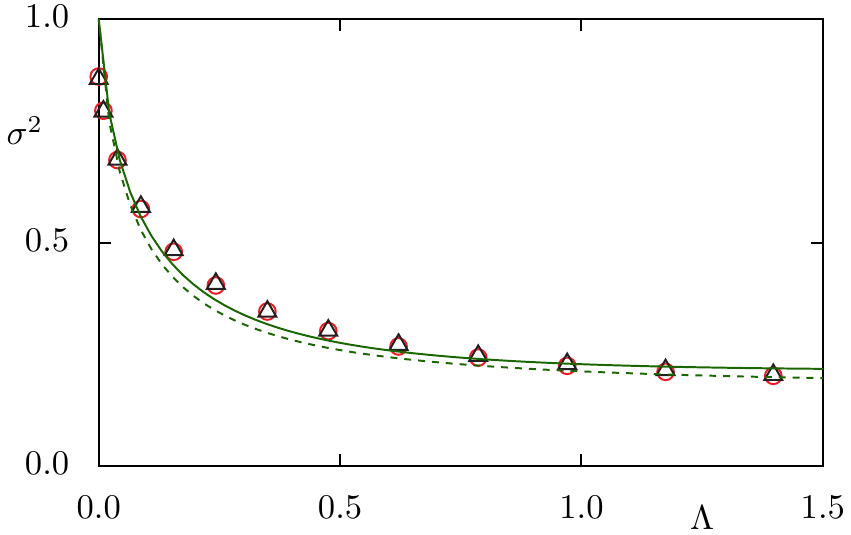}
\caption{(Color online)
  \label{fig:variance} Variance of the nearest neighbor spacing as
  function of the transition parameter. Shown are results for the
  coupled rotors (circles) and
  the RMT model (triangles), in both cases for $N=50$. The
  perturbation theory result is shown as solid line, Eq.~(\ref{sigmatwo}), while the dashed line
  refers to a $2\times 2$ transition ensemble result~\cite{Kota99}.}
\end{figure}

Thus if $\Delta E_i^0$ is the spacing between two neighboring levels at
($i,i+1$) before perturbation, it becomes $\Delta E_i \approx \Delta E_i^0+2
|V_{i,i+1}|^2/\Delta E_i^0$, ignoring the effect of other levels. Here
$V_{i,i+1}$ is an interaction matrix element, whose real and imaginary parts
are complex Gaussian random variables with a scale
determined by the variance $v^2$ introduced above.
This is appropriate given the normal
fluctuations of the real and imaginary parts of the GUE matrix
elements. Scaling the spacings and matrix elements, $r=\Delta E_i /D,s_0=\Delta E^0_i /D$ and $|V_{i,i+1}|^2=v^2w$, introduces $\Lambda$ as above, $r \approx s_0+2 \Lambda w /s_0$, with
$w$ and $s_0$ distributed exponentially, the latter due to the Poissonian
nature of the unperturbed spectrum.
Close lying or nearly degenerate levels are frequent for
an unperturbed Poissonian spectrum.  The resulting divergences are 
regularized using degenerate perturbation theory, which gives
\begin{equation}
\rho(r) = \int^\infty_0{\rm d}s_0\,  {\rm d}w \exp(-s_0-w) \delta \left(r-\sqrt{s_0^2+4\Lambda w}\right)
\end{equation}
The result of integration is
\begin{equation}
\label{NNS1}
\rho(r) = \frac{r}{\sqrt{\Lambda}}\left[ {\rm e}^{-r}D_+\left( \frac{r-2\Lambda}{2\sqrt{\Lambda}} \right) + {\rm e}^{-\frac{r^2}{4\Lambda}}D_+\left(\sqrt{\Lambda}\right) \right]
\end{equation}
where $D_+(x)$ is the Dawson function $ {\rm e}^{-x^2}\int_0^x {\rm d}t\ {\rm
  e}^{t^2}.$ The effect of all the other levels is accounted for by
recompressing the spectrum to unit mean spacing using
\begin{equation}
\label{NNSm}
\langle r \rangle = 1 + \Lambda  \left[ \frac{2}{\sqrt{\pi}} D_+\left( \sqrt{\Lambda} \right) - {\rm e}^{-\Lambda}{\rm Ei}\left( \Lambda \right) \right]
\end{equation}
where ${\rm Ei}(x)$ is the exponential integral.  Thus, the rescaled NNS distribution
is $P(s)= \langle r \rangle \rho(s \langle r \rangle )$. Also, the variance of the NNS is
\begin{equation}
\label{sigmatwo}
\sigma^2 = \frac{2+4\Lambda}{\langle r \rangle^2} - 1,
\end{equation}
and is a useful parameter to monitor the whole transition.
Figure~\ref{fig:n2b} compares the NNS obtained using
Eqs.~(\ref{NNS1}) and (\ref{NNSm}) with those of  the
coupled rotors, and is an excellent approximation.
Figure~\ref{fig:variance} compares the variance across the whole
transition.  While the perturbation theory result is slightly better than a
$2\times 2$ model results~\cite{Kota99} for $\Lambda \lesssim 1.0$,
the latter by design does better for larger $\Lambda$.
A more complete theory for the NNS covering the
whole transition is left for future work.

\emph{Summary and outlook.}---
In summary, this paper addresses an important, rather general
question of what happens when two chaotic systems, or systems with random
matrix fluctuations interact. The resulting entanglement leads to a universal 
transition and rapid recovery of global RMT fluctuations.  The
 transition is very well captured by a natural random matrix 
 ensemble, Eq.~(\ref{eq:mod}).  A universal scaling parameter is derived 
 in terms of the interaction strength and separately the abstract coupling 
 parameter of the RMT ensemble, thus establishing a quantitative relationship between the two objects.  Finally, perturbation theory with a simplified 
 global Poisson plus RMT ensemble reproduces the NNS transition in fluctuations very well.  In spite of its application to a particular dynamical
system, the coupled kicked rotors, the methods and results are of a general
nature, and can be expected to hold wherever there is a global coupling of two
subsystems each exhibiting RMT fluctuations.

Various extensions of this work are of significant interest. One rather
immediate generalization is to subsystems with different dimensions, say a
bipartite system with differing numbers of spins in the subsystems.  In many of
the relations, the substitution $N^2 \rightarrow N_1N_2$ suffices to capture
the new dependences.  A much more involved change is to a greater number of
subsystems.  If the number of subsystems is $L$, each with a dimensionality
$N$, a new expression for the transition parameter equation in
Eq.~(\ref{eq:lambda1}) is possible.  There, the new RMT model
analogous to Eq.~(\ref{eq:mod}) would have $L$ CUE matrices in a tensor product
and a global diagonal coupling of the same kind.  That leads to a
generalization of Eq.~(\ref{eq:lambdarmt}): $\Lambda=N^{2L}(1-\text{sinc}^2
(\pi \epsilon))/[4 \pi^2(N+1)^L] \approx \epsilon^2 N^L/12$.  Thus for large
$N$, it is expected that the recovery of RMT fluctuations is much faster with
increasing $L$.  However, this
analysis assumes essentially that $U_L(\epsilon)$ is an $L$-body operator.  If
it is restricted to a lower rank form, say a $2$-body operator, more
sophisticated models
have to be separately considered.  Preliminary results have also been obtained regarding
the statistics of eigenfunctions, and entanglement in the eigenstates which
develop solely due to the interactions.  They are beyond the scope of this
paper and left for future publication.

\bibliography{quantumchaos,rmtmodify,classicalchaos}

\end{document}